\documentclass[a4paper,12pt]{article}
\usepackage[left=2.5cm,bottom=3cm,right=2.5cm,top=3cm]{geometry} 

\usepackage{overpic,youngtab}
\usepackage{subfigure}
\usepackage[latin,english]{babel}
\usepackage{amsmath}
\usepackage{amssymb}
\usepackage{epstopdf}
\usepackage{graphics,psfrag,rotating}
\usepackage{mathabx}
\usepackage{graphicx}
\usepackage{dcolumn}
\usepackage{float}
\usepackage{pdflscape}
\usepackage{array}
\usepackage{booktabs}
\usepackage{amscd} 
\usepackage{mathtools}
\usepackage{fancybox}
\usepackage{fix-cm}
\usepackage[colorlinks=true,citecolor=blue,,linktocpage=true,linkcolor=blue,urlcolor=black]{hyperref}
\usepackage{tikz} 
\usetikzlibrary{matrix} 
\usetikzlibrary{positioning} 
\usetikzlibrary{arrows}
\usepackage{titlesec}
\usepackage{abstract}

\def\be{\begin{equation}}
\def\ee{\end{equation}}
\def\bea{\begin{eqnarray}}
\def\eea{\end{eqnarray}}

\def\a{\alpha}
\def\b{\beta}
\def\g{\gamma}
\def\d{\delta}
\def\m{\mu}
\def\n{\nu}
\def\t{\tau}

\def\l{\lambda}

\def\r{\rho}

\def\s{\sigma}

\def\bi{\begin{itemize}}
	\def\ei{\end{itemize}}

\begin{document}
	
	\vspace*{-1cm}
\phantom{hep-ph/***} 
{\flushleft
	{{FTUAM-23-xx}}
	\hfill{{ IFT-UAM/CSIC-23-133}}}
\vskip 1.5cm
\begin{center}
	{\LARGE\bfseries  The origin of the cosmological constant in unimodular gravity.}\\[3mm]
	\vskip .3cm
	
\end{center}

\vskip 0.5  cm
\begin{center}
	{\large Enrique \'Alvarez, Jes\'us Anero and Irene S\'anchez-Ruiz.}
	\\
	\vskip .7cm
	{
		Departamento de F\'isica Te\'orica and Instituto de F\'{\i}sica Te\'orica, 
		IFT-UAM/CSIC,\\
		Universidad Aut\'onoma de Madrid, Cantoblanco, 28049, Madrid, Spain\\
		\vskip .1cm

		\vskip .5cm
		
		\begin{minipage}[l]{.9\textwidth}
			\begin{center} 
				\textit{E-mail:} 
				\tt{enrique.alvarez@uam.es},
				\tt{jesusanero@gmail.com} and
				\tt{irenesanchezl2@gmail.com}
			\end{center}
		\end{minipage}
	}
\end{center}
\thispagestyle{empty}

\begin{abstract}
	\noindent It is well-known that in unimodular gravity (UG) the cosmological constant is not sourced by a constant energy density, but rather appears as some sort of integration constant. In this work we try to flesh this out by studying in some detail a couple of examples, one from cosmology and the other from gravitational collapse.
	
\end{abstract}

\newpage
\tableofcontents
\thispagestyle{empty}
\flushbottom

\newpage

\section{Introduction.}
Unimodular Gravity (UG) is a modification of General Relativity (GR) where only unimodular metrics (determinant, $g\equiv \left|det\,g_{\m\n}\right|=1$) are considered. Even in the path integral we are instructed to integrate over unimodular metrics only. The equations of motion  of UG are not identical to GR Einstein's ones with the same source. In fact the second Noether theorem (Bianchi identities)  allows to recover the lost trace as an integration constant. What happens then is that given a fixed value for the cosmological constant (CC) (for example $\Lambda=0$) any solution of GR is also a solution of UG, but the converse is untrue.  If we represent by $E$ the space of classical solutions of the equations of motion, then
\be
E_{GR (CC=\Lambda)}\subset E_{UG}
\ee 
and, somewhat symbolically,
\be
\sum_{\Lambda\in \mathbb{R}}  E_{GR (CC=\Lambda)}\equiv  E_{UG}
\ee 
From the physical point of view, UG is interesting because the value of the CC is {\em not} determined by the constant vacuum energy, which does not weigh at all in UG. The natural question is then, what determines the value of the CC in UG?.

In this paper we want to study this question under two different aspects, the first one dealing with  cosmological solutions and the second one with the spherical collapse.

To begin with, in \cite{Alvarez:2021cxy} we asked the same question in the framework of standard cosmology. We found there (and we review here) that in vacuum there are two solutions: one in which the scale factor is constant (which corresponds to flat space)
\be
b(t)=b_0
\ee
 and another  where
\be
b(t)=(3t-t_0)^{4/3}
\ee
In this case at least, the divide between these two solutions  lies in whether  the initial condition on the derivative of the scale factor $\dot{b}(t_0)$ is different from zero or not. This initial condition on a vacuum solution is admissible only in UG.
\par
Our second subject is the detailed study of gravitational collapse. This is of course a complicated aspect, but it is believed that all $j\ge 2$ multipoles of the matter are eventually radiated away, and that the final state is stationary, with axial symmetry \cite{Chrusciel}. In this paper we shall consider the simplest models of spherical collapse of a compact matter: the one originally proposed by Schwarzschild \cite{Schwarzschild}, where the source is assumed to be an incompressible perfect fluid (that is, with constant density)  and another where the source is assumed to be a presureless perfect fluid (a.k.a. dust) \cite{Weinberg:1972kfs}.
\par
What we find is that in GR the energy density of the collapsing cloud    $\r$ is bound to be related to the  curvature by 
\be \r=-{R\over \kappa^2}\ee 
owing to Einstein's equations. In UG this is not the case, and in fact 
\be \r=-{R\over \kappa^2}+\text{constant}\ee 
(note, however that the derivative $\dot{\rho}$ obeys the same restriction as in GR). In fact  this constant is intimately related to the value of the CC.
\section{Cosmological solutions}

The Friedmann-Lema\^itre metric  in the unimodular gauge of GR  (which is the only admissible metric in UG  cf. for example \cite{EAEVA})  reads
\be
ds^2=b(t)^{- 3/2}\,dt^2-b(t)^{1/2}\, \d_{ij}dx^i dx^j
\ee
where the function $b$ depends on time only, $b=b(t)$. The cosmic normalized four velocity vector field, $u^\m u_\m=1$, is given explicitly by
\be u^{\m}=\left(b^{3/4},0,0,0\right)\ee

It is easy to check that this  congruence is geodesic
\be
\dot{u}^\m\equiv \label{c}u^\n\nabla_\n u^\m=0
\ee
and the  expansion reads
\be \label{v1}\theta\equiv \nabla_\m u^\m=\frac{3}{4}b^{-1/4}\frac{db}{dt}\ee

\par
The equation of motion in  UG (the traceless piece of Einstein's) reads
\be 
R_{\m\n}-\dfrac{1}{4}R\,g_{\m\n}=\kappa^2 \left(T_{\m\n}-\dfrac{1}{4}\,T\,g_{\m\n}\right)\label{UGEM} 
\ee
We assume matter as a perfect fluid, that is
\be
T_{\m\n}\equiv \left(\r+p\right)u_\m u_\n-p g_{\m\n}
\ee
the energy-momentum conservation, $\nabla_\n T^{\m\n}=0$ is then equivalent to 
\be
\dot{\r}+\left(\r+p\right)\theta=0
\ee
Using the equation of motion \eqref{UGEM} and the property \eqref{c}, Raychaudhuri's equation \cite{Raychaudhuri:1953yv} reduces to
\be
\dot{\theta}+\frac{1}{n-1}\theta^2+\s_{\a\b}\s^{\a\b}-\omega_{\a\b}\omega^{\a\b}+\frac{1}{n}R+\frac{2(n-1)}{n}\kappa^2(\r+p)=0\ee
The scalar curvature reads
\be R=-2u^\m\nabla_\m{\theta}-\frac{4}{3}\theta^2\ee
Assuming as usual for simplicity vanishing shear and rotation, $\s_{\a\b}=\omega_{\a\b}=0$, and in the physical dimension $n=4$ 
\be\label{Re} \dot{\theta}+3\kappa^2(\r+p)=0\ee
It is worth remarking that it is not possible to express $R$ in terms of $T$.
\par
We can use   Ellis' clever trick \cite{Ellis}  to define a length scale through
\be \label{v2}\theta=3{\dot{ l}\over l}\ee
Then
\be
b\sim l^4.
\ee
Finally we can write  Raychaudhuri's equation \eqref{Re} as
\be\label{le} u^\m u^\n\Big[ l\nabla_\n\nabla_\m{l}-\nabla_\m l\nabla_\n l\Big]+\kappa^2(\r+p)l^2=0\ee
\subsection{Vacuum solutions.}
Vacuum implies   $p=\r=0$ and Raychaudhuri's equation reduces to
\be\label{J} \dot{\theta}=0\ee
In our case
\be \dot{\theta}=u^\n\nabla_\n\nabla_\m u^\m=-\frac{3}{16\sqrt{b}}\left[\left(\frac{db}{dt}\right)^2-4b\frac{d^2 b}{dt^2}\right]\ee
It follows that
\be\left(\frac{db}{dt}\right)^2-4b\frac{d^2 b}{dt^2}=0\ee

Its general solution is given by either a constant
\be
b=H_0^{4 \over 3}
\ee
 which corresponds to flat space; or else
\be 
b(t)=H_0^{4 \over 3}\left(3t-t_0\right)^{4\over 3}
\ee
which corresponds to  {\em de Sitter} \footnote{
	In unimodular coordinates, the maximally symmetric, constant curvature {\em de Sitter} spacetime reads
	\be\label{CC}
	ds^2=\left({dt\over 3 H t}\right)^2-(3 H t)^{2/3} \d_{ij}\,dx^i dx^j
	\ee
	that is, precisely
	\be
	b(t)\sim \,t^{4\over 3}
	\ee} space; $H_0\equiv 3\theta$ being the constant expansion. In this solution it is arbitrary, because there is no physical scale in the problem that determines it. It is to be emphasized that this solution depends on two parameters, whereas the  constant solution depends only on one, being thus less generic.
	\par
	This could be anticipated, because the vacuum equation of motion in unimodular gravity are just Einstein spaces 
	\be
	R_{\m\n}={1\over 4} R g_{\m\n}
	\ee
	Flat space is just a quite particular solution; constant curvature space-times \cite{Wolf} are a more generic one.
\subsection{The line $p+\r=C$.}

Let us examine the inhomogeneous equation of state $\r+p=C$. This is one of the most interesting results of UG, where physics depends on the value of the constant $C$ only.
\par
Indeed, depending on the value of the constant $C$, it is possible that both $p$ and $\r$ are positive. Only in the case $C=0$ is this situation strictly equivalent to a vacuum energy density.The differential equation \eqref{le} then reads
\be
\frac{1}{l}{d^2 l\over dt^2}-\frac{1}{l^2}\left({dl\over dt}\right)^2+C\kappa^2l^{-4}=0
\ee
whose general solution reads
\be l_C(t)=e^{-\sqrt{C_2}(t+C_3)}\Big\{\frac{1}{6C_2}\Big[e^{6\sqrt{C_2}(t+C_3)}-3C\kappa^2C_2\Big]\Big\}^{1/3}\ee
Obviously when $C=0$ this solution reduces to the vacuum solution
\be l_{\Lambda}(t)=l_0 e^{H_0t}\ee
with $C_3=0$ and $C_2^2=H_0$. 
\par
More interestingly, this solution is an atractor asymptotically when $t\rightarrow\infty$. Any solution tends asymptotically to de Sitter.
\par
For $C C_2 >0$ there is an {\em origin of time} $t_0$. For earlier times $t< t_0$ the solution becames unphysical. To be specific,
\be
t_0={1\over 6 \sqrt{C_2}}\log\,\left(3 C \kappa^2 C_2\right)- C_3
\ee
\subsection{Unimodular gravity versus General Relativity.}
The unimodular gauge of General Relativity (GR) is of course fully equivalent to the usual formulation of GR in comoving coordinates \cite{Bondi, Weinberg}  where the metric reads
\be\label{mc}
ds^2= d\t^2-a(\t)^2 \sum\d_{ij} dx^i dx^j
\ee
with a four velocity
\be u^\m=(1,0,0,0)\ee
and 
\be \dot{u}^\m=0\ee
so that in this case 
\be 
\theta=3\frac{1}{a}\frac{da}{dt}
\ee

We insist that he only difference between GR and UG stems from the equations of motion. Let us spell this out in some detail.

Now the equation of motion is the usual Einstein one
\be 
R_{\m\n}-\dfrac{1}{2}R\,g_{\m\n}=\kappa^2 T_{\m\n}\label{EM} 
\ee
in this case the scalar of curvature reads
\be R=-\frac{6}{a^2}\left[\left(\frac{da}{dt}\right)^2+a\frac{d^2a}{dt^2}\right]\ee
Then  Raychaudhuri's equation  in comoving coordinates yields
\be -3\left(\frac{da}{dt}\right)^2+2a^2\kappa^2\r=0\ee

In this case, the vacuum solution reduces to
\be \frac{da}{dt}=0\ee
i.e $\theta=0$ which is just flat spacetime, it is a subset of the UG result, $\dot{\theta}=0$ \eqref{J}, which is obviously a more general equation of motion.
\section{Spherical collapse}
We shall present two very simplified analysis. The first analysis  follows closely  the classic approach  pioneered by Schwarzschild \cite{Schwarzschild} and later by Oppenheimer and Volkoff \cite{Oppenheimer} for an incompressible fluid (which means constant density). 
The second one corresponds to what astrophysicists call dust (that is pressureless matter) cf. for example \cite{Weinberg:1972kfs}.
\par
The main idea is the following. It is assumed the existence of a radial coordinate, $r$, such that at a given value  $r=a$ a matching can be  performed between the {\em interior} solution sourced by the  energy momentum tensor of the perfect fluid, and an {\em exterior} solution, which according to Birkoff's theorem \cite{Birkhoff} must be Schwarzschild's.\footnote{
In fact a careful statement of Birkhoff's theorem \cite{Birkhoff} allows for Schwarzschild-de Sitter (SdS) or Schwarzschild-anti de Sitter (SadS) as exterior solutions.}
The first thing to note is that the interior solution is {\em not} Ricci flat (in fact the trace of the GR equations of motion implies that the curvature scalar is given by $R=-\kappa^2 T$, where $T$  is the trace of the energy-momentum tensor. For the equation of state this is just 
\be
R=-\kappa^2 \left(\r-3p\right).
\ee
This is at variance with the exterior solution, which is indeed Ricci flat according to Birkhoff's theorem. Then in the matching there is necessarily a discontinuity in the second derivatives of the metric.
\par
As we shall see in the next paragraph, this is not so in UG, where the interior metric is allowed to be Schwarzschild-(anti)-de Sitter, so that 
\be R=-\kappa^2\r-4 \Lambda\ee 
where $\Lambda$ is the cosmological constant. What happens is that the UG equations of motion are traceless, which loosens somewhat the constraint on the scalar curvature, allowing it to be modified by a constant term, precisely related to the CC. The main purpose of our work  is to understand better the physical meaning of the CC from the collapsing matter viewpoint.
\par
Let us first analyze some general properties of the energy-momentum tensor. For matter which can be modelled as a perfect fluid
\be
T_{\m\n}=(\r+p )u_\m u_\n-p g_{\m\n}
\ee
Covariant conservation of the energy-momentum tensor $\nabla_\n T^{\m\n}=0$ is equivalent to
\be
\nabla^\l p=\left(\dot{\r}+\dot{p}\right)u^\l+\left(\r+p\right)\left(\theta u^\l+\dot{u}^\l\right)
\ee
In the synchronous gauge this implies
\be
\dot{\r}+(\r+p)\theta=0
\ee
\subsection{A short review of the GR spherical collapse.}

\bi
\item When dealing with an  incompressible fluid, where $\r=\r_0$, the metric reads \cite{Schwarzschild}
\be 
ds^2=f_4dx_4^2-f_1dx_1^2-f_2\Big[\frac{dx_2^2}{1-x_2^2}+(1-x_2^2)dx_3^2\Big]
\ee
where the functions $f_i$ depend on $x_1$. If  we choose to work in  the unimodular gauge of GR (as Schwarzschild did)   we have to impose
\be f_4f_1f_2^2=1\ee

The interior solution turns out to be
\bea ds^2=\left(\frac{3\cos\chi_a-\cos\chi}{2}\right)^2dt^2-\frac{3}{\kappa^2\r_0+\Lambda}\left[d\chi^2+\sin^2\chi\Big[\frac{dx_2^2}{1-x_2^2}+(1-x_2^2)dx_3^2\Big]\right]\eea
whereas the exterior solution reads
\be ds^2=\Big[1-\frac{r_s}{\bar{r}}\Big]d\bar{t}^2-\frac{d\bar{r}^2}{\Big[1-\frac{r_s}{\bar{r}}\Big]}-\bar{r}^2\Big[\frac{dx_2^2}{1-x_2^2}+(1-x_2^2)dx_3^2\Big]\ee
Matching  both solutions, at the surface of the sphere $\bar{r}_a$, leads to
\be \cos^2\chi_a=1-\frac{r_s}{\bar{r}_a}\ee
where
\be \bar{r}_a=\sqrt{\frac{3}{\kappa^2\r_0}}\sin\chi_a\ee
The total mass of our sphere will be
\be M=\frac{3}{4\kappa^2}\sqrt{\frac{3}{\kappa^2\r_0}}\left(\chi_a-\frac{1}{2}\sin 2\chi_a\right)\ee
	
\item Assume  now that the collapsing cloud is made of  dust ($p=0$), with synchronous metric \cite{Weinberg:1972kfs}
\be ds^2=dt^2-f(r)S^2(t)dr^2-S^2(t)r^2d\Omega_2^2\ee

We have
\bea
&\theta=3 {\dot{S}\over S}\nonumber\\
&\dot{u}^\l=0
\eea

Conservation of the energy-momentum tensor implies that
\be\r(t)=\r(0)S^{-3}(t)\ee

The interior solution, valid for $r\le  a$, reads
\be ds^2=dt^2-S^2(t)\Big[\frac{dr^2}{1-\lambda r^2}+r^2d\Omega^2_2\Big]\ee
where $S(t)$ stands for
\bea &&S[\psi[t]]=\frac{1}{2}\left(1+\cos[\psi[t]]\right)\nonumber\\
&&t[\psi]=\left(\frac{\psi+\sin\psi}{2\sqrt{C}}\right)
\eea
with $C=\frac{\kappa^2\r_0}{3}$

The metric outside the sphere $r\ge a$ to be matched with this interior one must be the usual Schwarzschild metric owing to Birkhoff's theorem
\be ds^2=\Big[1-\frac{r_s}{\bar{r}}\Big]d\bar{t}^2-\frac{d\bar{r}^2}{\Big[1-\frac{r_s}{\bar{r}}\Big]}-\bar{r}^2d\bar{\Omega}^2\ee
 The actual matching of both solutions, in the surface of the sphere $\bar{r}=S(t)a$, leads to
\be M=\r_0\frac{4\pi}{3}a^3\ee

This metric can be easily written in the unimodular gauge by using  new coordinates $x$ and $\t$ such that  the dependence between new and old coordinates reads  $r(x)$ and $t(\t)$. Then $dr=r'(x)dx$ and $dt=t'(\t)d\t$ the metric becomes
\be ds^2=\left[t'(\t)\right]^2d\t^2-S^2[t(\t)]\Big[\frac{\left[r'(x)\right]^2}{1-\lambda r^2(x)}dx^2+r^2(x)d\Omega^2\Big]\ee
with the unimodular condition
\be\label{uc} 1=\left[t'(\t)\right]^2S^6[t(\t)]\frac{\left[r'(x)\right]^2}{1-\lambda r^2(x)} r^4(x)\ee
we split in two equations
\bea\label{UGcondition}
&&\left[t'(\t)\right]^2S^6[t(\t)]=C\nonumber\\
&&\frac{\left[r'(x)\right]^2}{1-\lambda r^2(x)} r^4(x)=\frac{1}{C}
\eea
Therefore the unimodular metric, with $C=1$, is
\be ds^2=\frac{d\t^2}{S^6[t(\t)]}-S^2[t(\t)]\Big[\frac{dx^2}{r^4(x)}+r^2(x)d\Omega^2\Big]\ee

It should be noted that the trace of Einstein's equations 
\be
R+\kappa^2 \r=0
\ee
is automatically enforced at all points, so the density at the surface (the point where the interior solution  should be matched with the exterior solution)  is not arbitrary, but determined by the geometry through the trace equation, as we have pointed out already.

\ei

\subsection{Unimodular presureless collapse.}
After this short review, let us repeat the analysis in UG using the same simplifying hypothesis on the matter source as before.

First, we assume $\r(r,t)$ and $p(r,t)=0$ and  the unimodular metric to be
\be \label{metric}ds^2=A^2(r,t)dt^2-B^2(r,t)dr^2-\frac{1}{A(r,t)B(r,t)}\Big[\frac{d\psi^2}{1-\psi^2}+(1-\psi^2)d\phi^2\Big]\ee
In this case, the conservation of energy-momentum tensor reduces to
\bea &&\nabla_\m T^{\m\n}=u^\n  u^\m\nabla_\m \r  +\r \left[u^\n\nabla_\m u^\m+u^\m\nabla_\m u^\n\right]\nonumber\\
\eea

 Bianchi identity yields
\be R+\kappa^2\r=\text{constant}
\ee
Later on, this constant will be identified with $-4\Lambda$.

Assume again a separable solution \cite{Weinberg}.
\bea
&&A(r,t)=\frac{1}{S^3[t(\t)]}\nonumber\\
&&B(r,t)=\frac{S[t(\t)]}{b^2(r)}
\eea
that is,  the metric is
\be ds^2=\frac{d\t^2}{S^6[t(\t)]}-\frac{S^2[t(\t)]}{b^4(r)}dr^2-b^2(r)S^2[t(\t)]\Big[\frac{d\psi^2}{1-\psi^2}+(1-\psi^2)d\phi^2\Big]\ee
If we combine the equation of motion and the Bianchi identity, we obtain
\be S^2[t(\t)]\Big\{ \Lambda+\frac{\kappa^2\r_0}{S^3[t(\t)]}-3S^4[t(\t)][S'[t(\t)]]^2\left[t'(\t)\right]^2\Big\}=\frac{3}{b^2(r)}\left[1-b^4(r)[b'(r)]^2\right]\ee
where we denote $S'[t(\t)]=\frac{S[t(\t)]}{dt}$ and the relation
\be \label{EQb}1+b^4(r)[b'(r)]^2+b^5(r)b''(r)=0\ee
From the unimodular condition \eqref{UGcondition} it  follows 
\be\label{EQS} \Lambda S^2[t]+\frac{\kappa^2\r_0}{S[t]}-3[S'[t]]^2=\kappa^2\r_0\ee
It is plain that  in the case $\Lambda=0$, we recover the GR solution.

It is easy to find a formal solution of 

\be\Lambda S^2[\psi[t]]+\frac{\kappa^2\r_0}{S[\psi[t]]}-3[S'[\psi[t]]]^2=\kappa^2\r_0\ee
namely
\bea &&S[\psi[t]]=\frac{1}{2}\left(1+\cos[\psi[t]]\right)
\eea
and
\be [\psi '[t]]^2=\frac{\kappa^2\r_0}{3}\sec^4\left[\frac{\psi[t]}{2}\right]+\frac{\Lambda}{3}\cot^2\left[\frac{\psi[t]}{2}\right]\ee
whose implicit solution reads
\bea t=\pm\sqrt{3}\int_1^{\psi}dx\frac{\sqrt{2+\cos[x]-2\cos[2x]-\cos[3x]}}{\sqrt{16\kappa^2\r_0\left(1-\cos[x]\right)+\Lambda\left(10+15\cos[x]+6 \cos[2x]+\cos[3x]\right)}}\eea

On the other hand, the proper energy density
\be \r=\frac{\r_0}{S^3[\psi[t]]}\ee
Therefore when $\psi=\pi$ the density $\r$ diverges.

\subsubsection{Expansion in  ${\Lambda\over \kappa^2 \r_0}$}
Let us expand  in ${\Lambda\over \kappa^2 \r_0}$ in order to get a grasp of the physical properties of our UG solution 
\be\label{EQSs}\Lambda S^2[t]+\frac{\kappa^2\r_0}{S[t]}-3[S'[t]]^2=\kappa^2\r_0\ee
with
\be S[t]=S_0[t]+\Lambda f[t]\ee
Introducing in \eqref{EQSs} the GR solution which is valid at order zero  in the expansion,
\bea
&&[S'_0[t]]^2=\frac{\kappa^2\r_0}{3}\Big[\frac{1}{S_0[t]}-1\Big]
\eea
At first  order, 
\be S_0^2[t]-\frac{\kappa^2\r_0 f[t]}{S_0^2[t]}-6f'[t]S_0'[t]=0\ee

Let us now perform the change  of variables
\be\label{1OL} S_0^2[\psi[t]]-\frac{\kappa^2\r_0 f[\psi[t]]}{S_0^2[\psi[t]]}-6f'[\psi[t]]S_0'[\psi[t]][\psi '[t]]^2=0\ee
in terms of the  the GR solution 
\bea \label{CYC} &&S_0[\psi[t]]=\frac{1}{2}\left(1+\cos[\psi[t]]\right)\nonumber\\
&&t[\psi]=\left(\frac{\psi+\sin\psi}{2\sqrt{\frac{\kappa^{2}\rho_{0}}{3}}}\right)
\eea

Substituting \eqref{CYC} into \eqref{1OL} we get:
\bea
\cos^8\left[\frac{\psi[t]}{2}\right]-\kappa^2\r_0\Big(f[\psi[t]]-\sin[\psi[t]]f'[\psi[t]]\Big)=0
\eea
where  $\psi[t]$   is given by \eqref{CYC}. Then
\be f[\psi[t]]=C_1\tan\left(\frac{\psi[t]}{2}\right)+\frac{560 t + 512 \cot(\tfrac{\psi[t]}{2}) + 376 \sin(\psi[t]) + 40 \sin(2 \psi[t]) + \tfrac{8}{3} \sin(3 \psi[t])}{512 \kappa^2 \rho_0 \cot(\tfrac{\psi[t]}{2})}\ee

The proper energy density is given by 
\be \r=\frac{\r_0}{S^3(\psi[t])}=\frac{\r_0}{\left(S_0(\psi[t])+\Lambda f(\psi[t])\right)^3}=\frac{\r_0}{S_0^3(\psi[t])}-\frac{3\r_0f(\psi[t])}{S_0^4(\psi[t])}\Lambda+\mathcal{O}(\Lambda^2)\ee
in such a way that when $\psi=\pi$, using \eqref{CYC}
\be T=\frac{\pi}{2}\sqrt{\frac{3}{\kappa^{2}\rho_{0}}}\ee
the density $\r$ diverges.

\subsubsection{Matching with the exterior solution.}
The generalized Birkhoff's theorem ensures that the metric outside the sphere $r=a$ must be the usual Schwarzschild-(anti)de Sitter 
\be ds^2=\Big[1-\frac{r_s}{\bar{r}}-\frac{\bar{r}^2\Lambda}{3}\Big]d\bar{t}^2-\frac{d\bar{r}^2}{\Big[1-\frac{r_s}{\bar{r}}-\frac{\bar{r}^2\Lambda}{3}\Big]}-\bar{r}^2d\bar{\Omega}^2\ee
Inside the sphere $r\le a$ we have
\be ds^2=\frac{d\t^2}{S^6[t(\t)]}-\frac{S^2[t(\t)]}{b^4(r)}dr^2-b^2(r)S^2[t(\t)]\Big[\frac{d\psi^2}{1-\psi^2}+(1-\psi^2)d\phi^2\Big]\ee
In order to match both solutions at the surface, is necessary to  redefine $\bar{r}=S[t(\t)]b(r)$. After some simple algebra, one gets

\be ds^2=\Big[1-\frac{\kappa^2\r_0}{3\bar{r}}b^3(a)-\frac{\bar{r}^2\Lambda}{3}\Big]d\bar{t}^2-\frac{d\bar{r}^2}{\Big[1-\frac{\kappa^2\r_0}{3\bar{r}}b^3(a)-\frac{\bar{r}^2\Lambda}{3}\Big]}-\bar{r}^2d\bar{\Omega}^2\ee
where  $r_s=2MG$ and the mass is given by
\be M=\r_0\frac{4\pi}{3} b^3(a)\ee

\subsection{UG  with incompressible fluid (constant density)} 
In this final section, we assume that the density is constant, $\r(r,t)=\r_0$ whereas the pressure is to be calculated $p(r,t)$. We further assume that the unimodular metric only depends on the radial coordinate
\be\label{metricAB}
ds^2= A^2(r) dt^2-B^2(r) dr^2-r^2 -{1\over A(r)B(r)}\Big[\frac{d\psi^2}{1-\psi^2}+(1-\psi^2)d\phi^2\Big]
\ee
In this case, the conservation of energy-momentum tensor reduces to
\bea &&\nabla_\m T^{\m\n}=u^\n  u^\m\nabla_\m p +\left(\r_0+p\right)u^\m\nabla_\m u^\n-\nabla^\n p =0\eea
because $\theta=0$. The timelike components  is identically satisfied. The radial component implies
\be \r_0+p(r,t)=\frac{\g}{A(r)}\ee
where $\g$ is a constant.

The equations of motion read
\be\hat{R}_{\m\n}-\frac{1}{4}g_{\m\n} \hat{R}=\kappa^2\left(T_{\m\n}-\frac{1}{4}g_{\m\n}T\right)\ee

Bianchi  identity implies
\be \hat{R}+\kappa^2\left(\r_0-3p(r)\right)=-C
\ee
Again, later on this constant will be identified with the CC, $C=4\Lambda$.


Assume now (cf. \cite{Schwarzschild})
\bea &&A^2[r]\equiv\zeta[r]\eta^{-1/3}[r]\nonumber\\
&&B^2[r]\equiv \frac{1}{\zeta[r]\eta[r]}
\eea
Then the equations of motion reduces to
\bea\label{1} 
&&2\zeta\eta''=-3\kappa^2\g\zeta^{-1/2}\eta^{1/6}\nonumber\\
&&\zeta''=-2\eta^{-\frac{5}{3}}+3\kappa^2\g\eta^{-\frac{5}{6}}\eta^{-\frac{1}{2}}+\frac{5}{3}\zeta\eta^{-1}\eta''\eea
and the Bianchi identity, with the expression of $\zeta''$ \eqref{1}
\bea 2\zeta\eta''+\zeta'\eta'=3\eta^{-2/3}-3\kappa^2\r_0-3\Lambda\eea

To summarize, we can express the interior solution like
\bea\label{solutionp} ds^2=\left(\frac{3\cos\chi_a-\cos\chi}{2}\right)^2dt^2-\frac{3}{\kappa^2\r_0+\Lambda}\left[d\chi^2+\sin^2\chi\left(\frac{d\psi^2}{1-\psi^2}+(1-\psi^2)d\phi^2\right)\right]\eea
where
\bea &&\eta\equiv\left(\frac{3}{\kappa^2\r_0+\Lambda}\right)^{3/2} \sin^3\chi\nonumber\\
&&\zeta=\sqrt{\frac{3}{\kappa^2\r_0+\Lambda}} \sin\chi\left(\frac{3\cos\chi_a-\cos\chi}{2}\right)^2\eea
It is plain that when $\Lambda=0$ we go back to Schwarzschild's spacetime.

\subsubsection{Matching with the exterior solution.}
The metric outside the sphere must still be the usual Schwarzschild-(anti)de Sitter 
\be ds^2=\Big[1-\frac{r_s}{\bar{r}}-\frac{\bar{r}^2\Lambda}{3}\Big]d\bar{t}^2-\frac{d\bar{r}^2}{\Big[1-\frac{r_s}{\bar{r}}-\frac{\bar{r}^2\Lambda}{3}\Big]}-\bar{r}^2d\bar{\Omega}^2\ee
Inside the sphere we have the solution \eqref{solutionp},
In order to match solutions at the surface, we need to redefine $\bar{r}^2=\frac{3}{\kappa^2\r_0+\Lambda}\sin^2\chi$, so that
\be 
ds^2=\left(\frac{3\cos\chi_a-\cos\chi}{2}\right)^2dt^2-\frac{1}{\cos^2\chi}d\bar{r}^2-\bar{r}^2\Big[\frac{d\psi^2}{1-\psi^2}+(1-\psi^2)d\phi^2\Big]
\ee
At the surface of the sphere $\bar{r}_a$
\be \cos^2\chi_a=1-\frac{r_s}{\bar{r}_a}-\frac{\bar{r}_a^2\Lambda}{3}\ee
where
\be \bar{r}_a=\sqrt{\frac{3}{\kappa^2\r_0+\Lambda}}\sin\chi_a\ee
\newpage
\section{Conclusions.}
In this paper we have examined the possible origin of the cosmological constant in the context of Unimodular Gravity (UG). 
The popular belief that UG is just equivalent to General Relativity (GR) in the unimodular gauge is unfortunately not correct, or at least in need of serious nuances.
\par
The main difference with GR lies in the equations of motion. In GR they read
\be
R_{\m\n}-{1\over 2} R g_{\m\n}=\kappa^2 T_{\m\n}
\ee
which in the free case reduces to 
\be
R_{\m\n}=0
\ee
In UG only the trace-free part of the equations of motion holds; that is 
\be
R_{\m\n}-{1\over 4} R g_{\m\n}=\kappa^2 \left(T_{\m\n}-{1\over 4} T g_{\m\n}\right)
\ee
which in the free case reduces to
\be
R_{\m\n}-{1\over 4} R g_{\m\n}=0
\ee
which allows for non-vanishing constant curvature solutions. This clearly shows that UG and GR are {\em not} equivalent, even in the unimodular gauge for the latter. 
\par
What is true instead, is that given a particular solution of UG there is some value of the cosmological constant (CC) such that this metric is a solution of the GR equations with this particular value of the CC. 
\par
 But the essential point we would like to make in this paper is that this value is {\em not} determined by the constant vacuum energy density (in case there is one such), but by boundary conditions in the equations of motion. Those are the ones we spelled out in this paper, both in the cosmological setting and also in some simplifyied models of spherical collapse. 
 \par
 Our hope is to have explained clearly that the boundary (or initial) conditions in the UG equations of motion that give rise to  a corresponding GR solution with nonvanishing CC are peculiar to UG; those particular   initial conditions would not be admissible in GR. There we would need to put a CC by hand. 
 For example, the initial condition on the cosmological scale factor
 \be
 b(t=0)\neq 0
 \ee
 is only admissible in UG.
 The same thing happens with the initial condition
 \be
 R+\kappa^2\r\neq 0
 \ee
 in spherical gravitational collapse. In GR the second member is only allowed to vanish, whereas in UG it can have any real value.
\section{Acknowledgements}
We have enjoyed discussions with Eduardo Velasco-Aja. We acknowledge partial financial support by the Spanish MINECO through the Centro de excelencia Severo Ochoa Program  under Grant CEX2020-001007-S  funded by MCIN/AEI/10.13039/501100011033.
 
We also acknowledge partial financial support by the Spanish Research Agency (Agencia Estatal de Investigaci\'on) through the grant PID2022-137127NB-I00 funded by 
MCIN/AEI/10.13039/501100011033/ FEDER, UE

	All authors acknowledge the European Union's Horizon 2020 research and innovation programme under the Marie Sklodowska-Curie grant agreement No 860881-HIDDeN and also byGrant PID2019-108892RB-I00 funded by MCIN/AEI/ 10.13039/501100011033 and by ``ERDF A way of making Europe''.

\newpage

\newpage
  

\begin{thebibliography}{99}
	\bibitem{Alvarez:2021cxy}
E.~Alvarez and J.~Anero,
``Unimodular Cosmological models,''
[arXiv:2109.08077 [gr-qc]].
 \bibitem{EAEVA}
E.~Alvarez and E.~Velasco-Aja,
``A Primer on Unimodular Gravity,''
[arXiv:2301.07641 [gr-qc]].
\bibitem{Birkhoff}
Birkhoff, G. D. (1923),
" Relativity and Modern Physics".
 Cambridge, Massachusetts: Harvard University Press. LCCN 23008297.
 \bibitem{Bondi}
 H. Bondi, Spherically Symmetrical Models in General Relativity, Monthly Notices of the Royal Astronomical Society, Volume 107, Issue 5-6, December 1947, Pages 410–425, https://doi.org/10.1093/mnras/107.5-6.410
	\bibitem{Chrusciel}
P.~T.~Chrusciel, J.~Lopes Costa and M.~Heusler,
``Stationary Black Holes: Uniqueness and Beyond,''
Living Rev. Rel. \textbf{15} (2012), 7
doi:10.12942/lrr-2012-7
[arXiv:1205.6112 [gr-qc]].
 \bibitem{Ellis}
Ellis, G. F. R.,
``Relativistic cosmology,''
Proc. Int. Sch. Phys. Fermi \textbf{47} (1971), 104-182
doi:10.1007/s10714-009-0760-7
	
	\bibitem{Oppenheimer}
J.~R.~Oppenheimer and G.~M.~Volkoff,
``On massive neutron cores,''
Phys. Rev. \textbf{55} (1939), 374-381
doi:10.1103/PhysRev.55.374
\bibitem{Raychaudhuri:1953yv}
A.~Raychaudhuri,
Phys. Rev. \textbf{98} (1955), 1123-1126
doi:10.1103/PhysRev.98.1123
	\bibitem{Schwarzschild}
K.~Schwarzschild,
``On the gravitational field of a sphere of incompressible fluid according to Einstein's theory,''
Sitzungsber. Preuss. Akad. Wiss. Berlin (Math. Phys. ) \textbf{1916} (1916), 424-434
[arXiv:physics/9912033 [physics.hist-ph]].
\bibitem{Weinberg}
S.~Weinberg  (2008) ,
``Cosmology,''
Oxford Univ. Pr.
S.~Weinberg,
``The Cosmological Constant Problem,''
Rev. Mod. Phys. \textbf{61}, 1-23 (1989)
doi:10.1103/RevModPhys.61.1
	\bibitem{Weinberg:1972kfs}
	S.~Weinberg,
	``Gravitation and Cosmology: Principles and Applications of the General Theory of Relativity,''
	John Wiley and Sons, 1972,
	ISBN 978-0-471-92567-5, 978-0-471-92567-5
	\bibitem{Wolf}
	Wolf, J. A. (1967). ``Spaces of constant curvature''. New York: McGraw-Hill.\\
	Petrov, A. Z. (1969). ``Einstein spaces ''. Oxford: Pergamon Press.\\
	Besse, A. L. (1987). ``Einstein manifolds''. Berlin: Springer-Verlag.
	
\end{thebibliography}
\end{document}